\documentstyle[prd,aps,preprint,tighten,pictex]{revtex}

\begin{document}

\draft
\preprint{\begin{tabular}{r}
KIAS-P99043 \\
hep-ph/9906392
\end{tabular}}

\title{Nonthermal Axino as Cool Dark Matter\\ in Supersymmetric Standard Model
       without R-parity}
\author{
    Eung Jin Chun$^{a*}$ 
and Hang Bae Kim$^{b\dagger}$ 
}
\address{
    $^a$Korea Institute for Advanced Study \\
    207-43 Cheongryangri-dong, Dongdaemun-gu,
    Seoul 130-012, Korea \\
    $^b$Department of Physics and Institute of Basic Science\\
    Sungkyunkwan University, Suwon 440--746, Korea\\
    $^*$ejchun@kias.re.kr,
    $^\dagger$hbkim@newton.skku.ac.kr
}
\maketitle

\begin{abstract}
We point out that the axino predicted as the supersymmetric partner of 
the axion is a good candidate for the recently proposed sterile neutrino 
cool dark matter.  
The axino mass falls into the right range ($\lesssim$ keV) 
in the context of gauge mediated supersymmetry breaking.  
A sizable mixing of the axino with active neutrinos arises
when R-parity violation is allowed and the resulting neutrino 
masses and mixing accommodate the atmospheric and solar 
neutrino oscillations simultaneously.
\end{abstract}



In a recent paper \cite{cldm}, a new cosmological model has been proposed to
explain the structure formation of the Universe.  In this model, dark matter
consists of non-thermal sterile neutrinos with masses about 100 eV to 
10 keV which are produced around the big bang nucleosynthesis (BBN) 
epoch through a resonant active-sterile neutrino conversion in the presence
of a net lepton number asymmetry.
Such non-thermal  neutrinos are called ``cool'' since their free streaming
scale is larger than that of the usual cold dark matter but smaller 
than that of hot or warm dark matter.  This cool dark matter (ClDM) model
would be a good alternative to the conventional cold plus hot dark matter
(CHDM) model which has been known to be the best cosmological model for the 
structure formation \cite{chdm}.  In view of particle physics theory, 
the validity of the CHDM model depends crucially on  the origin of
active neutrino masses and mixing.
In the CHDM model, 20 \% of dark matter is composed of hot components
which are unarguably taken as three species of active neutrinos.
The cosmological requirement,  $\Omega_\nu \simeq 0.2$, constrains
the sum of three active neutrino masses through the relation,
$\sum m_\nu = 92 \Omega_\nu h^2$ eV, which implies
$\sum m_\nu \simeq 4.6$ eV, taking $h=0.5$.  
If neutrinos take natural hierarchical masses as quarks and charged leptons,
the recent atmospheric neutrino data from the Super-Kamiokande \cite{skatm} 
imply $\Delta m^2_{\rm atm} \approx m^2_{\nu_3} < 10^{-2}$ eV$^2$
and thus $\sum m_\nu < 0.1 $ eV, in which case the CHDM model cannot work.
%

The purpose of this paper is to present a particle physics model in which
the ClDM model is realized in a natural way while neutrinos obtain 
hierarchical masses which can accommodate the atmospheric \cite{skatm} and 
solar neutrino data \cite{bks} simultaneously.   One of popular ways to
generate neutrino masses is to consider the minimal supersymmetric extension
of standard model without imposing R-parity in which R-parity and lepton number
violating bilinear and trilinear couplings give rise to
typically hierarchical neutrino masses \cite{rpnms,newrp}.
If one introduces in this context the Peccei-Quinn (PQ) mechanism which
would be the most attractive solution to the strong CP problem \cite{jekim},
there exits a singlet fermion, called the axino $\tilde{a}$, 
which is a supersymmetric
partner of the Goldstone boson, the axion, of a spontaneously broken
PQ symmetry.  
Various cosmological and astrophysical observations are known to 
restrict the scale of PQ symmetry breaking:  $f_a\approx
(10^{10}-10^{12})$ GeV \cite{kt}.  As was observed first in Ref.~\cite{cjs},
the axino can be a sterile neutrino, that is, it mixes with active
neutrinos once R-parity is not imposed.
In the context of supergravity  \cite{sugra}
where supersymmetry breaking is mediated
at the Planck scale, the axino mass is quite model-dependent and
could be in the range, $100 \,{\rm GeV} -1 \,{\rm keV}$ \cite{gy}.  However, 
in models with gauge mediated supersymmetry breaking (GMSB) \cite{gmsb} where
the supersymmetry breaking mediation scale is much below the PQ scale $f_a$, 
the axino mass is determined by the Goldstone nature of the axion 
supermultiplet, and is predicted to be typically in the sub-keV regime 
in the minimal gauge mediation models \cite{chun}.

We will show that the axino can be a good candidate of cool dark matter 
as its mass and mixing with active neutrinos fall into the right ranges 
for a reasonable range of parameter space in the context under consideration.
In particular, the required mixing of the axino with active neutrinos can
be obtained when one allows the R-parity violating terms whose sizes are
fixed to explain the atmospheric and solar neutrino masses and mixing.
A weak point of ClDM scenario would be the requirement of a large lepton 
asymmetry. It will be argued that the demanded lepton asymmetry can 
be generated through a late-time entropy production followed by the
R-parity and lepton number violating decays
of the ordinary lightest supersymmetric particle (LSP), 
or through the Affleck-Dine mechanism.

\medskip

There are various sources for the cosmological axino population.
The conventional source would be the thermal production.  If the reheat
temperature $T_R$ after inflation is larger than the axino decoupling 
temperature $T_d \sim f_a$, thermally produced axinos may overclose the universe
unless the axino mass is less than about 2 keV \cite{rtw}.  Therefore,
the axino with mass $\sim$ keV can be {\it warm} dark matter.  
When $T_R < T_d$, the primordial axinos are inflated away,
but can be regenerated from thermal background.  In this case,
the axino {\it warm} dark matter needs the  mass \cite{chang};
\begin{equation} \label{regen}
 m_{\tilde{a}} \approx 3\,{\rm MeV} \left(f_a \over 10^{12}\,{\rm GeV}\right)^2
                  \left(10^9\,{\rm GeV} \over T_R \right) \,.
\end{equation}
The other possibility is the secondary production of the axinos through 
the decay of ordinary superparticles.  In a recent paper \cite{covi}, it 
was argued that the axino with mass $\sim$ 10 GeV can be {\it cold}
 dark matter when the axinos are produced from the decay of 
the ordinary neutralino LSP.
The last one would be the nonthermal production resulting in the axino 
{\it cool} dark matter, which we would like to discuss here.

Let us  recapitulate the basic ingredients 
of the ClDM model realizing the nonthermal production \cite{cldm}.
The production of a sterile neutrino $\nu_s$ ($=\tilde{a}$ in our scheme)
which is much heavier than
active neutrinos can be made by a resonant active-sterile neutrino conversion
driven by a pre-existing large lepton asymmetry $L=10^{-4}-10^{-1}$
which is destroyed during this process.
The resonant oscillation between active and sterile neutrinos occurs at a
temperature
\begin{equation}
T_{\rm res} \approx 9 \left(\frac{m_{\nu_s}}{100\,{\rm eV}}\right)^{1/2}
    \left(\frac{{\cal L}_i}{0.1}\right)^{-1/4} 
    \left(\frac{E}{T}\right)^{-1/4}\,{\rm MeV},
\end{equation}
where ${\cal L}_i=2L_{\nu_i}+\sum_{j\ne i}L_{\nu_j}$.
The resonant transformation is adiabatic when the active-sterile neutrino
vacuum mixing is not too small, that is, 
\begin{equation} \label{thetasi}
\sin^22\theta_{si}\gtrsim10^{-9}
\end{equation}
for the lepton asymmetry $L=10^{-4}-10^{-1}$.
Since the resonance temperature of a low energy neutrino is higher than that
of a high energy neutrino, low energy neutrinos are produced at first
consuming most of the lepton asymmetry.  This makes high energy neutrino
conversion non-adiabatic and no significant conversion occurs for high energy
neutrinos.  The produced sterile neutrinos are thereby cooler
than the active neutrinos and the spectrum is non-thermal
with $\langle E\rangle/T\approx0.7$.  The free-streaming length of these
nonthermal sterile neutrinos is
\begin{equation} \label{fs}
\lambda_{\rm fs} \sim \left(\frac{270\,{\rm eV}}{m_{\nu_s}}\right)
    \left(\frac{\langle E\rangle/T}{0.7}\right)\, {\rm Mpc}\,,
\end{equation}
and the contribution to the matter density today is given by
\begin{equation}  \label{Os}
\Omega_{\nu_s} \approx \left(\frac{m_{\nu_s}}{343\,{\rm eV}}\right)
    \left(\frac{0.5}{h}\right)^2 \left(\frac{{\cal L}_i}{0.1}\right) \,.
\end{equation}
In this mechanism, the sterile neutrino production can be confidently
calculated only at the temperatures below the quark-hadron phase
transition temperature, about 150 MeV.  This translates to the condition
\begin{equation} \label{qh}
 m_{\nu_s} < 23 \left(\frac{{\cal L}_i}{0.1}\right)^{1/2} {\rm keV} \,.
\end{equation}
{}From Eqs.~(\ref{Os})--(\ref{qh}), one concludes that about 100 eV to 
10 keV sterile neutrinos produced through the active--sterile neutrino 
conversion driven by a lepton asymmetry $L \approx10^{-4}-10^{-1}$ 
can be a ClDM candidate.

\medskip

Our framework is the minimally extended supersymmetric standard model (MSSM)
allowing R-parity and lepton number violating interactions which can explain
the neutrino masses and mixing implied by the atmospheric and solar neutrino
data.  In order to estimate  the axino mass and mixing with active neutrinos,
we need to know the sizes of R-parity violating parameters determined
from the neutrino data \cite{skatm,bks}.
The R-parity and lepton number violating terms in the MSSM superpotential are
\begin{equation} \label{Wrp}
 W=\epsilon_i \mu  L_i H_2 + 
      \lambda_{ijk} L_i L_j E^c_k + \lambda'_{ijk} L_i Q_j D^c_k \,,
\end{equation}
where $\mu$ is the Higgs mass parameter in the R-parity conserving 
superpotential, $W=\mu H_1 H_2$.  
As is well known \cite{rpnms}, (active) neutrinos get masses at tree level
due to the sneutrino vacuum expectation values (VEVs), $\langle L^0_i \rangle$,
misaligned with the bilinear terms $\epsilon_i$, as well as
at 1-loop level through squark and slepton/Higgs exchanges. 
The tree-level neutrino mass matrix takes the form \cite{jaja};
\begin{equation} \label{mtree}
 m^{\rm tree}_{ij} \approx \xi_i \xi_j {M_Z^2 \over M_{1/2}} c^2_\beta \,,
\end{equation}
where $M_Z$ is the Z boson mass, $M_{1/2}$ is the typical gaugino (photino 
or zino) mass, and $c_\beta \equiv\cos\beta$. Note that  
$t_\beta\equiv\tan\beta=\langle H_2 \rangle / \langle H_1 \rangle$. 
Here $\xi_i \equiv \langle L_i^0\rangle /\langle H_1 \rangle -\epsilon_i$ 
measure the misalignment between the sneutrino VEVs and $\epsilon_i$.
Only one neutrino obtains a nonzero mass from the tree contribution 
(\ref{mtree}) and  the corresponding eigenvalue is given by
$m_{\nu_3}\approx\xi^2 (M_Z^2/M_{1/2})c^2_\beta$ where
$\xi \equiv \sqrt{\sum\xi_i^2}$.  
Now, the atmospheric neutrino data from Super-Kamiokande \cite{skatm}
imply that  $m_{\nu_3} \approx \sqrt{\Delta m^2_{\rm atm}} \approx 0.05$ eV and
$\xi_2 \approx \xi_3$ resulting in large mixing between the muon and
tau neutrino.  Therefore, one finds $\xi \sim \xi_{2,3}$ and 
\begin{equation} \label{xicos}
 \xi c_\beta \approx 10^{-6} \left(M_{1/2} \over M_Z\right)^{1/2}
                   \left(m_{\nu_3} \over 0.05\,{\rm eV}\right)^{1/2} \,.
\end{equation}
It is important for the later use to recall that nonzero values of $\xi_i$ 
arise through renormalization group evolution from the mediation scale $M_m$
to the weak scale and thus can be much smaller  than the original 
parameters $\epsilon_i$ or $\lambda, \lambda'$ in Eq.~(\ref{Wrp}). 
In one step approximation for integrating the 
renormalization group equation (See, for instances, Ref.~\cite{jv}.),
one obtains 
\begin{equation}
 \xi_i \sim \epsilon_i \left( \mu A_b \over m^2_{\tilde{l}} \right)
       \left( {3h^2_b \over 8\pi^2} \ln{M_m \over m_{\tilde{l}}} \right)
\end{equation}
where $h_b, A_b$ are the bottom quark Yukawa coupling and the corresponding 
soft-parameter respectively, and $m_{\tilde{l}}$ is the slepton mass.
Taking $\mu A_b =m^2_{\tilde{l}} and  M_m=10^3 m_{\tilde{l}}$,
one gets
\begin{equation} \label{epsi}
 \epsilon_i \sim 10^{-2}  c_\beta 
                \left( \xi_i c_\beta \over 10^{-6}\right) \,.
\end{equation}
The 1-loop contributions give nonzero masses to the other two
neutrinos, which are much smaller than the tree contribution,
in particular,  in the context of GMSB models \cite{hwang}.
Among various 1-loop contributions to the second largest neutrino mass 
$m_{\nu_2}$, the diagrams with interchange of slepton and charged 
Higgs become important for large $t_\beta$ and thus relevant for generating
the solar neutrino  mass scale.  According to the estimation 
in Ref.~\cite{hwang}, the mass ratio $m_{\nu_2}/m_{\nu_3}$ is given by
\begin{equation} \label{m2}
 \frac{m_{\nu_2}}{m_{\nu_3}} \sim 10^{-2}
          \left(\frac{\lambda_{233}}{\lambda'_{333}}\right)
          \left(\frac{t_\beta}{50}\right)^2 
\end{equation}
under the assumption of the usual hierarchy among the trilinear couplings,
that is, those involving third generations are larger than the others.
The solar neutrino data explained by the matter resonant conversion, or 
the vacuum oscillation \cite{bks} requires  $m_{\nu_2}  \approx 
\sqrt{\Delta m^2_{\rm sol}} \sim 10^{-3}$, or $10^{-5}$ eV, respectively.
Therefore, the matter conversion  or vacuum oscillation solution 
can be obtained for $t_\beta \sim 50$ or 10, respectively,
assuming $\lambda_{233} \approx \lambda'_{333}$. 
Note also that one needs $\lambda_{233}, \lambda'_{233, 333} \sim
(10^{-4}-10^{-6}) (m_{\nu_2}/10^{-3} \,{\rm eV})^{1/2}$ for the right
solar neutrino mass scale \cite{newrp,hwang}.

\medskip

Let us now turn to the question why the axino can be ClDM 
in the context of the R-parity violating MSSM with gauge mediated 
supersymmetry breaking.  
As discussed in Ref.~\cite{chun}, if a superfield $S$ carries a 
PQ charge $x_S$, there arises the following effective K\"ahler potential 
for the superfield $S$ and the axion superfield $\Phi$, 
which nonlinearly realizes the spontaneously broken PQ symmetry 
below its breaking scale $f_a$,
\begin{equation} \label{Keff}
 K= {x_S \over f_a} (\Phi^\dagger + \Phi) S^\dagger S \,.
\end{equation}
Upon supersymmetry breaking, the interaction in Eq.~(\ref{Keff}) 
gives rise to a mass mixing between the axino 
and the fermionic component of $S$ given by 
\begin{equation} \label{mD}
 m_D \approx x_S {F_S \over f_a}
\end{equation}
where $F_S$ is the F-term of the field $S$.  Without fine-tuning of 
some parameters, the value of $F_S$ is expected to be of the order of
$M_S^2$ when the field $S$ has the mass $M_S$.  As a consequence, the axino
gets a see-saw suppressed mass, $ m_{\tilde{a}} \approx m_D^2/M_S  
\sim x_S^2 M_S^3/f_a^2$, that is, 
\begin{equation} \label{mns}
 m_{\tilde{a}} \sim 10\,{\rm keV} \left(M_S \over 10^5 \,{\rm GeV} \right)^3
                             \left(10^{10}\,{\rm GeV} \over f_a \right)^2 \,.
\end{equation}
This shows that the axino has a mass in the right range for being a 
ClDM sterile neutrino, given that $f_a=(10^{10}-10^{12})$ GeV and  
$M_S=(10^5-10^7)$ GeV when $S$ is a field in a 
messenger (or a hidden) sector of minimal gauge mediation models 
\cite{mgm}.  

A sizable mixing of the axino with active neutrinos arises
due to the bilinear terms in the superpotential (\ref{Wrp}) 
when the lepton doublets are charged under the PQ symmetry.  
Similarly to Eq.~(\ref{mD}), the mixing mass between the axino and
the active neutrino $\nu_i$ is given by $m_{si} \approx x_L F_L/f_a$
where $x_L$ is the PQ charge of lepton doublet $L$ and
$F_L\equiv\epsilon_i\mu\langle H_2\rangle$.  Therefore,
using the estimation of $\epsilon_i$ in Eq.~(\ref{epsi}), one finds
\begin{equation} \label{msi}
 {m_{si}\over {\rm keV}} \sim 0.1 c_\beta s_\beta
   \left( \mu \over 500\,{\rm GeV} \right)
   \left(10^{10} \,{\rm GeV} \over f_a\right)
   \left( \xi_i c_\beta \over 10^{-6} \right) \,.
\end{equation}
Given $m_{\tilde{a}}$, the adiabaticity condition
$\theta_{si}\approx m_{si}/m_{\tilde{a}}\gtrsim10^{-5}$ (\ref{thetasi})
puts a lower bound on $m_{si}$,
and an upper bound comes from the fact that the see-saw reduced mass of
active neutrinos $m_{si}^2/m_{\tilde{a}}$ should be smaller than the value
$m_{\nu_3} \sim 0.05$ eV used in Eq.~(\ref{xicos}).
Taking roughly $m_{si}^2/m_{\tilde{a}}< 10^{-2}$ eV, we get the appropriate
range of mixing mass
\begin{equation} \label{msibound}
    10^{-2}\left(\frac{m_{\tilde{a}}}{{\rm keV}}\right)\,{\rm eV}
    \lesssim m_{si} \lesssim 
    3\left(\frac{m_{\tilde{a}}}{{\rm keV}}\right)^{1/2}\,{\rm eV} \,.
\end{equation}
{}From Eqs.~(\ref{mns}) and (\ref{msi}), one can find that the sterile
and active neutrino mixing mass can be obtained in a reasonable range of 
the parameter space under consideration.
%

We have now to comment on the lifetime of our sterile neutrino, the axino, 
which should be stable as a dark matter. 
As we are working with gauge mediated supersymmetry breaking which has
low supersymmetry breaking scale,
the axino may be able to decay into a gravitino and an axion.
For the supersymmetry breaking scale of order $ \sqrt{F} \lesssim 
10^6\,{\rm GeV}$ the gravitino mass is $m_{3/2} \lesssim 1$ keV
and thus the axino decay can be allowed kinematically.
In this case, the lifetime of the axino  is found to be \cite{hbkim}
\begin{equation}
\tau_{\tilde{a}} \approx 10^{29} \,{\rm sec}
    \left(\frac{1\,{\rm keV}}{m_{\tilde{a}}}\right)^5
    \left(\frac{\sqrt{F}}{10^5\,{\rm GeV}}\right)^4 \,.
\end{equation}
That is, the axino is stable in the cosmological time scale for 
all the parameter ranges under consideration.

\medskip

A nontrivial requirement for the ClDM scenario is a large lepton 
asymmetry, $L=10^{-4}-10^{-1}$, around the nucleosynthesis period.
This number is hierarchically larger than the baryon asymmetry
$B\simeq10^{-10}$ required for producing the observed light element
abundances through nucleosynthesis.
Therefore, in the absence of a cancellation between $L_{\nu_i}$,
the large lepton asymmetry must be generated at the temperature below
the electroweak phase transition temperature to avoid the transfer between
the lepton asymmetry and the baryon asymmetry due to the sphaleron effects.
In our framework, since neutrino masses come from L violating couplings,
they could also be the source of lepton asymmetry.
Then the most promising way of generating such a lepton asymmetry would 
be the Dimopoulos-Hall mechanism \cite{dh}.

When a late-time entropy production like thermal inflation 
is introduced for various cosmological reasons \cite{lyth},
the lepton asymmetry can be generated through the late-time decay of 
a (thermal) inflaton followed by lepton number violating superparticle decays 
\cite{dh}.  The lepton asymmetry after the reheat of a late-time thermal
inflation is then given by 
\begin{equation}
L \simeq 5\,\varepsilon_L\,\frac{T_R}{m_\phi}
\end{equation}
where $m_\phi$ is the mass of a thermal inflaton $\phi$,
$T_R$ is the reheat temperature after the decay of $\phi$, and 
$\varepsilon_L$ is the amount of lepton asymmetry per $\phi$ decay.
As a possibility to get a large $\varepsilon_L$, let us consider the case 
where the stau $\tilde{\tau}$ is the ordinary LSP which occurs 
in a wide range of the GMSB parameter space \cite{gmsb}.  
As a stau is the LSP among the ordinary sparticles, it may decay into
a tau and a gravitino ($\tilde{\tau}\to \tau \tilde{G}$), 
or into a neutrino and a charged lepton ($\tilde{\tau}\to \nu l$) through
$\lambda$ couplings in Eq.~(\ref{Wrp}). As we want to have  a large
lepton asymmetry, the latter decay rate has to be larger than  the former.
In other words, we require
$\Gamma(\tilde{\tau}\to\nu_l l)\approx m_{\tilde{\tau}}|\lambda|^2/16\pi >
 \Gamma(\tilde{\tau}\to\tau\tilde{G})\approx m_{\tilde{\tau}}^5/16\pi F^2$,
which puts a bound on $\lambda$ depending on the supersymmetry breaking
scale $\sqrt{F}$ as follows:
\begin{equation} \label{ldecay}
 \lambda > \frac{m_{\tilde{\tau}}^2}{F} \sim 10^{-6} 
 \left(\frac{m_{\tilde{\tau}}}{100\,{\rm GeV}}\right)^2
 \left(\frac{10^5\,{\rm GeV}}{\sqrt{F}}\right)^2 \,.
\end{equation}
In view of generating the solar neutrino mass scale  $m_{\nu_2}$ 
as given  in Eq.~(\ref{m2}) and the following discussions,  
the above inequality can be fulfilled for reasonable values of $\sqrt{F}$.
The argument goes parallel with the stau decay into two quarks through 
$\lambda'$ couplings which we neglect without loss of generality.
As far as Eq.~(\ref{ldecay}) is satisfied, the stau mainly decays 
through R-parity and lepton number violating interactions.
Out-of-equilibrium condition can be met if the reheat
temperature is smaller than $m_{\tilde\tau}/20$ which is the typical 
supersymmetric particle decoupling temperature \cite{kt}.
CP violation can come from complex phases of
the leptonic charged current in case of Majorana neutrino masses,
or from the complex L violating trilinear couplings $\lambda,\lambda'$.
Let us now assume for simplicity that the right-handed stau $\tilde{\tau}_R$
is the LSP which has decay modes $\tilde{\tau}_R \to \nu_i e_j$.
Via the tree and one-loop interference depicted in FIG.~1, it gives
\begin{equation} \label{epL}
\varepsilon_L \approx \frac{2g^2_2{\rm Im}\left(
    \sum\lambda_{ij3}\lambda^*_{lm3}U^*_{il}U_{mj}\right)}
    {4\pi\sum|\lambda_{ij3}|^2}
\end{equation}
where $U_{ij}$ is the CKM type mixing element for the charged leptonic current.
The factor 2 comes from the fact that each $\phi$ produces at least two staus.
When for instance $\lambda_{233}$ is the largest coupling,
Eq.~(\ref{epL}) becomes
\begin{equation}
    \varepsilon_L \approx 2\alpha_2 {\rm Im}(U^*_{22}U_{33}) \,.
\end{equation}
Note that $U^*_{22}U_{33}$ is complex in general for Majorana neutrinos
and its phase unconstrained by neutrino oscillation experiments
can be of order one.
Then the large mixing for atmospheric neutrino oscillations \cite{skatm}
implies ${\rm Im}(U^*_{22} U_{33})\approx 1/2$.
Taking the upper limit of $T_R\approx m_{\tilde{\tau}}/20$
which is required for a stau to be out of equilibrium, one finds
\begin{equation}
   L \approx \frac{\alpha_2}{4}\,\frac{m_{\tilde\tau}}{m_\phi} \,.
\end{equation}
Therefore, the maximal value of lepton asymmetry in this mechanism is 
$L\sim10^{-2}$.
For $L\sim10^{-3}$, Eq.~(\ref{qh}) implies $m_{\tilde{a}}\lesssim2$ keV.
{}From Eq.~(\ref{Os}), we obtain $\Omega_{\tilde{a}}\approx0.2$
for $m_{\tilde{a}}\approx2$ keV and $L=10^{-3}$.

An even larger lepton asymmetry can arise through the Affleck-Dine
mechanism \cite{ad} with a  low reheat temperature.
There is an interesting class of models where the lepton asymmetry $L$
of order 1 can be produced through the Affleck-Dine mechanism.
These are supergravity models that possess a so-called Heisenberg symmetry,
which include no-scale type supergravity
models and many string based supergravity models.
In these models, supersymmetry breaking by the inflationary vacuum energy
does not lift flat directions at tree level, and one-loop corrections
gives small negative mass square for flat directions not involving large
Yukawa couplings.  After inflation these flat directions generate a large
baryon and/or lepton asymmetry, typically of order 1,
through the Affleck-Dine mechanism \cite{gmo}.
For example, we can use the well-known $LH_2$ flat direction for this purpose,
with the lepton number violation and CP violation arising from
nonrenormalizable interactions.
In this mechanism, it is possible to obtain $L\approx10^{-1}$ and hence
$\Omega_{\tilde{a}}\approx1$.
One difficulty is that the large lepton asymmetry may accompany with
the large baryon asymmetry.
Several ways to reduce such large baryon asymmetry to the observed level
were discussed in Ref.~\cite{cgmo}.
As we require the reheat temperature $T_R$ to be below the electroweak scale,
the decay produced, or the regenerated axino population can be completely
negligible as can be seen from  Eq.~(\ref{regen}).

\medskip

In conclusion, we have presented a well-motivated candidate for 
nonthermal sterile neutrino dark matter.  The PQ mechanism realized in
the supersymmetric standard model to resolve the strong CP problem predicts
the presence of the axino, the fermionic superpartner of the axion.
The lightness of the axino can arise when supersymmetry breaking is
mediated by gauge interactions.
We have pointed out that the axino mass can fall naturally into the demanded 
range $(0.1-10)$ keV,  given the range of the supersymmetry breaking scale,
$M_S\approx (10^4-10^7)$ GeV, and the PQ symmetry breaking scale,
$f_a\approx(10^{10}-10^{12})$ GeV.
It has also been shown that a sizable mixing between active neutrinos and
the axino, $\theta>10^{-5}$, required for it to be sterile neutrino dark
matter, can arise from the (bilinear) R-parity violation which is
introduced to explain the atmospheric and solar neutrino masses and mixing.
An `ad hoc' feature of nonthermal sterile neutrino dark matter scenario is
that it needs a large lepton asymmetry $L\approx10^{-4}-10^{-1}$.
We have argued that the lepton asymmetry up to $L\sim 10^{-2}$ can be
obtained by a late-time (thermal) inflaton decay followed by R-parity
violating decays of the ordinary LSP $\tilde{\tau}$, 
given the maximal CP violating phases in the Majorana neutrino masses.
An even larger lepton asymmetry can result from the Affleck-Dine mechanism.
The reheat temperature is then required to be below the electroweak 
scale or a few GeV.
The scenario under consideration needs better understanding 
why there is a big hierarchy between baryon and lepton asymmetries.


\begin{figure}
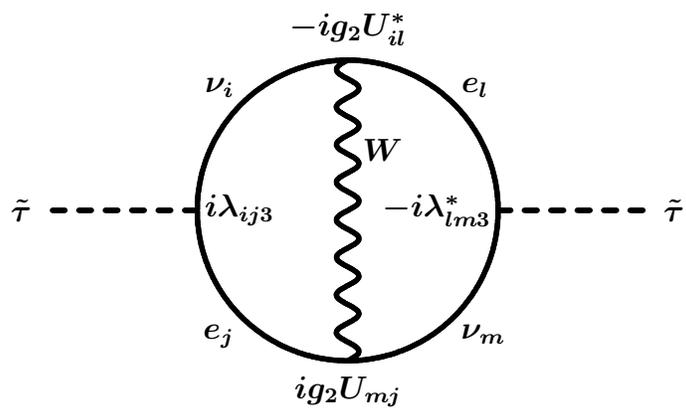

\begin{center}
\mbox{\beginpicture
\setcoordinatesystem units <10mm,10mm>
\setplotarea x from -4 to 4, y from -2 to 2
\unitlength=10mm
\linethickness=1pt
\put {\boldmath$i\lambda_{ij3}$}    [l] < 1mm,0mm> at -2  0
\put {\boldmath$-i\lambda_{lm3}^*$} [r] <-1mm,0mm> at  2  0
\put {\boldmath$ig_2U_{mj}$}    [t] <0mm,-2mm> at  0 -2
\put {\boldmath$-ig_2U_{il}^*$} [b] <0mm, 2mm> at  0  2
\put {\boldmath$\tilde\tau$} [r] <-2mm,0mm> at -4  0
\put {\boldmath$\tilde\tau$} [l] < 2mm,0mm> at  4  0
\put {\boldmath$e_l$}   [lb] < 1mm, 1mm> at  1.414  1.414
\put {\boldmath$\nu_i$} [rb] <-1mm, 1mm> at -1.414  1.414
\put {\boldmath$e_j$}   [rt] <-1mm,-1mm> at -1.414 -1.414
\put {\boldmath$\nu_m$} [lt] < 1mm,-1mm> at  1.414 -1.414
\put {\boldmath$W$} [l] <2mm,-2mm> at  0 1
\setplotsymbol ({\huge.})
\setlinear
\ellipticalarc axes ratio  1:1  360 degrees from 2 0 center at 0 0
\setdashes <5pt>
\plot -2 0  -4 0 /
\plot  2 0   4 0 /
\setsolid
\setquadratic
\plot
 0.000  -2.000
 0.037  -1.980
 0.072  -1.960
 0.103  -1.940
 0.127  -1.920
 0.143  -1.900
 0.150  -1.880
 0.147  -1.860
 0.136  -1.840
 0.116  -1.820
 0.088  -1.800
 0.055  -1.780
 0.019  -1.760
-0.019  -1.740
-0.055  -1.720
-0.088  -1.700
-0.116  -1.680
-0.136  -1.660
-0.147  -1.640
-0.150  -1.620
-0.143  -1.600
-0.127  -1.580
-0.103  -1.560
-0.072  -1.540
-0.037  -1.520
 0.000  -1.500
 0.037  -1.480
 0.072  -1.460
 0.103  -1.440
 0.127  -1.420
 0.143  -1.400
 0.150  -1.380
 0.147  -1.360
 0.136  -1.340
 0.116  -1.320
 0.088  -1.300
 0.055  -1.280
 0.019  -1.260
-0.019  -1.240
-0.055  -1.220
-0.088  -1.200
-0.116  -1.180
-0.136  -1.160
-0.147  -1.140
-0.150  -1.120
-0.143  -1.100
-0.127  -1.080
-0.103  -1.060
-0.072  -1.040
-0.037  -1.020
 0.000  -1.000
 0.037  -0.980
 0.072  -0.960
 0.103  -0.940
 0.127  -0.920
 0.143  -0.900
 0.150  -0.880
 0.147  -0.860
 0.136  -0.840
 0.116  -0.820
 0.088  -0.800
 0.055  -0.780
 0.019  -0.760
-0.019  -0.740
-0.055  -0.720
-0.088  -0.700
-0.116  -0.680
-0.136  -0.660
-0.147  -0.640
-0.150  -0.620
-0.143  -0.600
-0.127  -0.580
-0.103  -0.560
-0.072  -0.540
-0.037  -0.520
 0.000  -0.500
 0.037  -0.480
 0.072  -0.460
 0.103  -0.440
 0.127  -0.420
 0.143  -0.400
 0.150  -0.380
 0.147  -0.360
 0.136  -0.340
 0.116  -0.320
 0.088  -0.300
 0.055  -0.280
 0.019  -0.260
-0.019  -0.240
-0.055  -0.220
-0.088  -0.200
-0.116  -0.180
-0.136  -0.160
-0.147  -0.140
-0.150  -0.120
-0.143  -0.100
-0.127  -0.080
-0.103  -0.060
-0.072  -0.040
-0.037  -0.020
 0.000   0.000
 0.037   0.020
 0.072   0.040
 0.103   0.060
 0.127   0.080
 0.143   0.100
 0.150   0.120
 0.147   0.140
 0.136   0.160
 0.116   0.180
 0.088   0.200
 0.055   0.220
 0.019   0.240
-0.019   0.260
-0.055   0.280
-0.088   0.300
-0.116   0.320
-0.136   0.340
-0.147   0.360
-0.150   0.380
-0.143   0.400
-0.127   0.420
-0.103   0.440
-0.072   0.460
-0.037   0.480
-0.000   0.500
 0.037   0.520
 0.072   0.540
 0.103   0.560
 0.127   0.580
 0.143   0.600
 0.150   0.620
 0.147   0.640
 0.136   0.660
 0.116   0.680
 0.088   0.700
 0.055   0.720
 0.019   0.740
-0.019   0.760
-0.055   0.780
-0.088   0.800
-0.116   0.820
-0.136   0.840
-0.147   0.860
-0.150   0.880
-0.143   0.900
-0.127   0.920
-0.103   0.940
-0.072   0.960
-0.037   0.980
-0.000   1.000
 0.037   1.020
 0.072   1.040
 0.103   1.060
 0.127   1.080
 0.143   1.100
 0.150   1.120
 0.147   1.140
 0.136   1.160
 0.116   1.180
 0.088   1.200
 0.055   1.220
 0.019   1.240
-0.019   1.260
-0.055   1.280
-0.088   1.300
-0.116   1.320
-0.136   1.340
-0.147   1.360
-0.150   1.380
-0.143   1.400
-0.127   1.420
-0.103   1.440
-0.072   1.460
-0.037   1.480
-0.000   1.500
 0.037   1.520
 0.072   1.540
 0.103   1.560
 0.127   1.580
 0.143   1.600
 0.150   1.620
 0.147   1.640
 0.136   1.660
 0.116   1.680
 0.088   1.700
 0.055   1.720
 0.019   1.740
-0.019   1.760
-0.055   1.780
-0.088   1.800
-0.116   1.820
-0.136   1.840
-0.147   1.860
-0.150   1.880
-0.143   1.900
-0.127   1.920
-0.103   1.940
-0.072   1.960
-0.037   1.980
-0.000   2.000
/
\endpicture}
\end{center}
\caption{The diagram for tree and one-loop interference}
\end{figure}

\end{document}